\documentclass[prl,twocolumn,showpacs]{revtex4}
\usepackage{amsmath,amstext,amsthm,amssymb,amsfonts}
\usepackage{graphicx}

\begin{document}

\title{Dilepton yields from Brown-Rho scaled vector mesons including
memory effects}

\author{Bj\"orn Schenke}
\author{Carsten Greiner}

\affiliation{Institut f\"ur Theoretische Physik, %
   Johann Wolfgang Goethe -- Universit\"at Frankfurt, %
   Max--von--Laue--Stra\ss{}e~1, %
   D--60438 Frankfurt am Main, %
   Germany}

\begin{abstract}
Brown-Rho scaling, which has been strongly discussed after recent
NA60 data was presented, is investigated within a nonequilibrium
field theoretical description that includes quantum mechanical
memory. Dimuon yields are calculated by application of a model for
the fireball, and strong modifications are found in the comparison to
quasi-equilibrium calculations, which assume instantaneous
adjustment of all meson properties to the surrounding medium. In
addition, we show results for the situation of very broad
excitations.
\end{abstract}



\maketitle


 Motivated by the recent discussion on whether Brown-Rho scaling
 \cite{br91} is `ruled out' by the NA60 data, presented at Quark Matter 2005 \cite{Damjanovic:2005ni,Arnaldi:2006jq},
 we calculate dilepton yields from $\rho$-mesons with dropping masses within the nonequilibrium
 quantum field formulation introduced in \cite{Schenke:2005ry}.
 First quantitative but simplified calculations within this work
 have shown that memory effects, which are neglected in
 quasi-equilibrium calculations \cite{Rapp:1999ej} and
 \cite{Skokov:2005ut,vanHees:2006ng,vanHees:2006iv,Renk:2006ax},
 have an important influence
 on the final dilepton yields, especially for dropping mass scenarios.
 This lead to the conclusion that an analysis of which scenario describes the data correctly demands
 the proper inclusion of memory effects.
 The aim of this work is to clarify how the inclusion of memory
 affects the dimuon yields from Brown-Rho scaled vector mesons. The validity
 of standard equilibrium calculations depends on the strength of these effects.

 We consider two mass parameterizations: The
 temperature and density dependent one used by Rapp et al. \cite{rapp1}
 \begin{equation}
    m_{\rho}^{*}=m_{\rho}(1-0.15\,
    n_{\text{B}}/n_0)\left[1-(T/T_{\text{c}})^2\right]^{0.3}\text{,} \label{rapppar}
 \end{equation}
 assuming a constant gauge coupling $g$ in the vector meson dominance (VMD) -coupling,
 and one motivated by the renewed version of Brown-Rho scaling
 discussed in \cite{Brown:2005ka,Brown:2005kb}:
 \begin{equation}
    m_{\rho}^{*}=m_{\rho}(1-0.15\,n_{\text{B}}/n_0)\text{,}\label{brownpar}
 \end{equation}
 with a modified gauge coupling $g^*$, in such a way that $g^*$ is
 constant up to normal nuclear density $n_0$, while from then on
 $m^*_{\rho}/g^*$ is taken to be constant \cite{Brown:2002is}.
 This parameterization is not valid close to $T_c$ because it does
 not take the meson mass to zero at the critical point.
 However, Brown and Rho argue \cite{Brown:2005ka,Brown:2005kb} that
 because lattice calculations show that
 the pole mass of the vector mesons does not change
 appreciably up to $T=125$ MeV, the parameterization using
 $[1-(T/T_{\text{c}})^n]^{d}$ (with positive $d$ and integer $n$)
 overestimates the mass shift. They find temperature dependent effects to be
 an order of magnitude smaller than the density dependent
 effects, and hence suggest to concentrate on the density
 dependent part. We also wish to study the effect of finite memory
 for a mass shift following this parameterization and ignore
 its shortcomings for the moment.
 Furthermore, Brown and Rho point out that due to the
 violation of VMD, which accounts for most of the
 shape of the dilepton spectrum (see \cite{Brown:2005ka,Brown:2005kb} and
 \cite{Harada:2003jx}), the overall dilepton production in dense matter
 should be reduced by a factor of 4 compared to Rapp's
 calculations, which we do not take into account here.

 We compare the usual approach \footnote{We
 will refer to this as the Markovian approach, because memory is completely neglected.}, where the dilepton
 rate is given by the well known equilibrium formula
\begin{align}
    \frac{dN}{d^4x d^4k}(\tau,k)=&\frac{2 e^4}{(2\pi)^5}\frac{m_{\rho}^4}{g_{\rho}^2}\mathcal{L}(M)\notag\\
    &\times\frac{1}{M^2}n_{\text{B}}\left(T(\tau),k_0\right)\pi A_{\rho}
    (\tau,k)\text{,}
    \label{markovrate}
\end{align}
 with invariant mass $M^2=k^2=k_0^2-\mathbf{k}^2$ and $\mathcal{L}(M)=\left(1+\frac{2 m_l^2}{M^2}\right)\sqrt{1-\frac{4 m_l^2}{M^2}}\,\theta(M^2-4m_l^2)$,
 to the nonequilibrium formalism, in which the propagators of the
 $\rho$-meson and the virtual photon are calculated using the
 general nonequilibrium formulas. The dilepton rate for a spatially homogeneous but time dependent system is given
 by \cite{Schenke:2005ry}:
         \begin{align}
           \frac{dN}{d^4xd^4k}(\tau,k)=&\frac{2\,e^2}{(2\pi)^5}M^2\mathcal{L}(M)\notag\\
           &\times \Re\left[\int_{t_0}^{\tau}d\bar{t}\,i\,D_{\gamma\,T}^{<}(\textbf{k},\tau,\bar{t})e^{i
           k_0(\tau-\bar{t})}\right]\label{photrate1}\text{.}
        \end{align}
In its derivation we treated transverse and longitudinal modes
equally, which is adequate for our purposes. $D_{\gamma\,T}^{<}$
is the transverse virtual photon propagator, and satisfies the
generalized fluctuation dissipation relation
         \begin{equation}
            D^{<}=D^{\text{ret}} q^{<}D^{\text{adv}}\label{conv0}\text{,}
         \end{equation}
 with all time variables and integrations, and further indices implicit.
 The relation follows directly from the Kadanoff-Baym equations
 when terms corresponding to initial conditions at time $t_0$ are
 omitted. This can be done when the system is given enough time to
 reach its initial state before we start the actual
 calculation of the yield \cite{Schenke:2005ry}.
 In this case the
 $D^{\text{ret/adv}}$ are the free retarded and advanced virtual photon
 propagators while $q^<=\Pi^<$, the photon self energy,
 follows VMD by
\begin{equation}
            \Pi^<_{T}(t_1,t_2)=e^2\frac{m_{\rho}^{*\,2}(t_1)}{g_\rho^*(t_1)} D_{\rho\,T}^{<}(t_1,t_2)\frac  {m_{\rho}^{*\,2}(t_2)}{g_\rho^*(t_2)}\text{.}\label{photvmd}
\end{equation}
 Note that the couplings are to be taken at the correct vertices,
 separated in time. The free photon propagator has to be exponentially damped in
 order to reduce large contributions from early times, which
 make the extraction of the higher frequency structure numerically
 problematic \footnote{Explicitly, we use the replacement $D_{\gamma\,T}^{\text{ret}}(\tau-t_1) = (\tau-t_1) \rightarrow (\tau-t_1) e^{-\Lambda
 (\tau-t_1)}$ and an analogous one for the advanced propagator. Then we renormalize the result by multiplication with
 $(\omega^2+\Lambda^2)^2/\omega^4$ (see \cite{Schenke:2005ry} for more
 details).
 Note that a variation of $\Lambda$ between $200$ and
 $800$ MeV leads to a change in the resulting yield of up to 5\% in the interesting region between 0.4 and 0.8 GeV.
 Differences very close to twice the muon mass reach about 15\%.}.

 The meson propagator $D_{\rho\,T}^{<}$ also satisfies (\ref{conv0}),
 with $q^<=\Sigma^<$, the meson self energy, and the
 retarded and advanced meson propagators, which obey the
 equation of motion that in a spatially homogeneous and
 isotropic medium is given by
         \begin{align}
            \left(-\partial_{t_1}^2\right.&\left.-m_{\rho}^{*\,2}-\textbf{k}^2\right)D_{\rho\,T}^{\text{ret}}(\textbf{k},t_1,t_2)\notag\\
            &-\int_{t_2}^{t_1}d\bar{t}
            \Sigma^{\text{ret}}_{\rho\,T}(\textbf{k},t_1,\bar{t})D_{\rho\,T}^{\text{ret}}(\textbf{k},\bar{t},t_2)=\delta(t_1-t_2)\text{,}\label{photdgl}
        \end{align}
 and
 $D_{\rho\,T}^{\text{adv}}(\textbf{k},t_1,t_2)=D_{\rho\,T}^{\text{ret}}(\textbf{k},t_2,t_1)$.
 For calculating $\Sigma^<$ from $\Sigma^{\text{ret}}$ we assume local equilibrium and
 introduce the
 fireball temperature to the calculation.
 The integral over past times in Eq.
 (\ref{photrate1}) as well as the time integrations in Eq.
 (\ref{conv0}) have encoded the finite memory of the system.
 In \cite{Schenke:2005ry} we show that in order to describe
 a heavy ion reaction it is essential to retain this
 dynamic information. This is due to the
 time scales for the adaption of the meson's spectral properties to
 the evolving medium being of the same order as the lifetime
 of the regarded hadronic system. In this situation the evolution
 is non-Markovian and quantum mechanical interferences among past time contributions become
 important \cite{Schenke:2005ry}.
 Therefore a gradient approximation
 can not properly describe the situation in a heavy ion collision
 and remarkable differences between the dynamic and the Markovian yields occur.

\begin{figure}[tbh]
      \begin{center}
        \includegraphics[height=5.5cm]{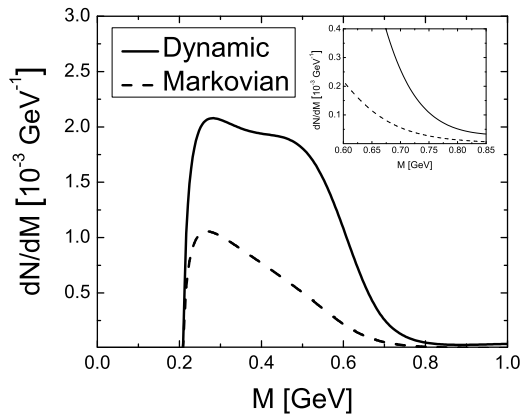}
        \caption{Dimuon yield from decaying $\rho$-mesons calculated out of equilibrium and within the usual static approximation using parameterization (\ref{rapppar}).
        A difference of about a factor of 4 is visible in the regime between 400 and 800 MeV.}
        \label{fig:rapp}
      \end{center}
\end{figure}

\begin{figure}[tbh]
      \begin{center}
        \includegraphics[height=5.5cm]{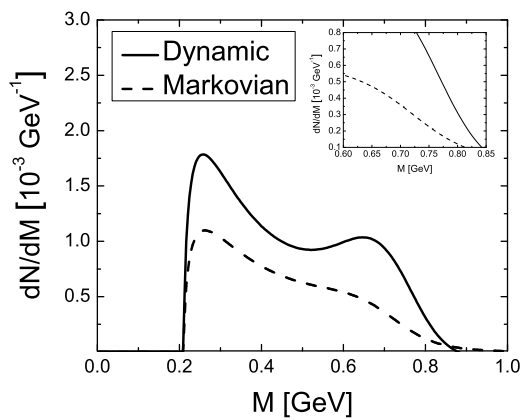}
        \caption{Dimuon yield from decaying $\rho$-mesons calculated out of equilibrium and within the usual static approximation using parameterization (\ref{brownpar}).
        The most significant difference (about a factor of 2-3) is visible between 600 and 800 MeV.}
        \label{fig:br}
      \end{center}
\end{figure}

    Following \cite{Rapp:1999us,vanHees:2006ng,vanHees:2006iv},
    we use a fireball model with a cylindrical volume expansion
    in the $\pm z$ direction:
    \begin{equation}
        V(\tau)=(z_0+v_z \tau)\pi(r_0+0.5\,a_{\perp}\tau^2)^2\text{,}
    \end{equation}
    where $z_0$ is equivalent to a formation time, $v_z=c$ is the
    primordial longitudinal motion, $r_0=5.15\,\text{fm}$ is the initial nuclear overlap radius,
    and $a_{\perp}=0.08\,c^2/\text{fm}$ is the radial acceleration.
    We start the calculation at $T=T_c=175$ MeV (at $\tau_0\approx 3$ fm/c) and set the meson self energy
    $\Sigma^<$ to zero for temperatures below chemical freezeout at $T=120$ MeV, thus ending the production of $\rho$-mesons.
    The hadronic phase, for which we calculate the dimuon production, lives for about $6$ fm/c.
    The time dependent temperature and density, determined via the fireball
    evolution, yield the time dependence of the $\rho$-meson mass
    following parameterizations (\ref{rapppar}) and (\ref{brownpar}).
    We include additional broadening by about 100 MeV as it was
    done in Rapp's calculation \cite{rapp1} shown in \cite{Arnaldi:2006jq}.
    In the dynamic case the $\rho$-meson propagator and the photon
    propagator follow Eq.\,(\ref{conv0}) after a Fourier
    transformation of the self energy into the two time representation.
    The rate is then calculated using Eq.\,(\ref{photrate1}),
    whereas in the quasi-equilibrium case the rate at a certain time follows
    directly from Eq.\,(\ref{markovrate}), using the spectral
    function and Bose-distribution corresponding to the present mass and
    temperature.
    The calculation is done for each momentum mode between $k=0$ and $k=1.5$ GeV and integrated over momenta.
    We show the comparison of the dynamic and the
    quasi-equilibrium calculation for the two parameterizations in Figs. \ref{fig:rapp} and
    \ref{fig:br}.
    We find a strong enhancement as well as modification of shape in both cases. For the density and temperature
    dependent parameterization the dynamic calculation yields about a
    factor of 4 more dileptons between invariant masses of 400 and 800 MeV as compared to the Markovian case, while the
    other parameterization shows an enhancement of about 2-3 times between 600 and 800 MeV in the dynamic calculation.

    The reasons for the strong differences are the following: First,
    the meson mass effectively approaches its vacuum value more slowly
    in the dynamic calculation. Lower masses are enhanced due to the
    Bose factor and if the spectral function has a lower mass for a
    longer time enhancement is greater. Second, the
    memory of higher temperatures increases this enhancement for all
    masses. Finally, the modified coupling in Eq. (\ref{photvmd})
    suppresses the early contributions. This means that for the static
    case all low masses in the spectral function and high temperatures
    are suppressed. However, due to the finite memory in the dynamic case, these low masses and
    high temperatures still contribute at later times when the coupling is
    larger, meaning that they are less suppressed by the VMD-coupling.

    In the first case, besides the overall increase in
    yield, we find a significant enhancement in the
    dynamic calculation for invariant masses around the
    $\rho$-vacuum mass. Comparing to NA60 data (shown vs. the equilibrium Brown-Rho calculation in e.g.
    \cite{Brown:2005ka,Arnaldi:2006jq}), the enhancement would improve the situation for the
    Brown-Rho scenario, which in the equilibrium calculation
    strongly underestimates the data in this regime. The second
    parameterization leads to even more weight around the vacuum
    peak in the dynamic case and would also improve the agreement
    with the data.

\begin{figure}[tbh]
      \begin{center}
        \includegraphics[height=5.5cm]{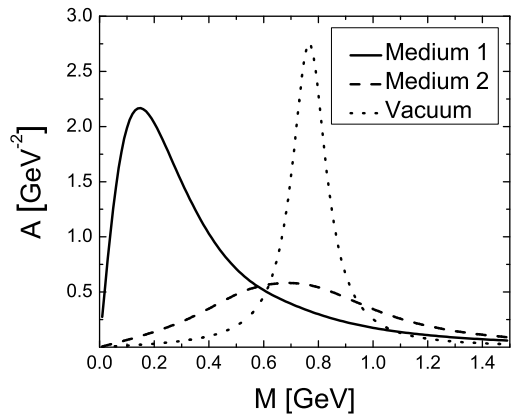}
        \caption{Medium modified spectral function using self energy (\ref{modelshenself}) with (Medium 1: $m=300$ MeV) and without (Medium 2) mass shift at normal nuclear
        density and widths of $\Gamma=700$ MeV and $\Gamma_R=400$ MeV. The $\rho$-vacuum spectral function is shown for comparison.}
        \label{fig:spectral}
      \end{center}
\end{figure}

Having investigated pure dropping mass scenarios, we now combine
Brown-Rho-scaling using parameterization (\ref{rapppar}) with
strong broadening and coupling to the excitation $N^*(1520)$. This
is achieved by using the self energy \cite{Schenke:2005ry}
        \begin{align}
            \text{Im}\Sigma^{\text{ret}}&(\tau,\omega,\textbf{k})=-\frac{\rho(\tau)}{3}
            \left(\frac{f_{RN\rho}}{m_{\rho}}\right)^2\notag\\
            &\times g_I \frac{\omega^3 \bar{E}\,\Gamma_R(\tau)}{(\omega^2-\frac{\Gamma_R(\tau)^2}{4}-\bar{E}^2)^2+(\Gamma_R(\tau)\omega)^2}-\omega\Gamma(\tau)
            \label{modelshenself}\text{\,,}
        \end{align}
with $\bar{E}=\sqrt{m_R^2+\textbf{k}^2}-m_N$ and $m_R$ and $m_N$
the masses of the resonance and nucleon, respectively \cite{po04}.
$\Gamma_R$ is the width of the resonance, which in vacuum is 120
MeV, and $g_I$ the isospin factor. Both widths $\Gamma$ and
$\Gamma_R$ increase substantially in the medium
($\Gamma\rightarrow 700$ MeV and $\Gamma_R\rightarrow 400$ MeV)
\cite{vanHees:2006ng,Rapp:1997ei}. The transverse and longitudinal
components of the self energy (\ref{modelshenself}) have been
combined, following our approximation of treating both
contributions equally. Note that we are always using the vacuum
$\rho$-mass in Eq. (\ref{modelshenself}) to prevent the coupling
to the resonance from becoming infinite. Typical resulting
spectral functions are shown in Fig. \ref{fig:spectral}.

\begin{figure}[tbh]
      \begin{center}
        \includegraphics[height=5.5cm]{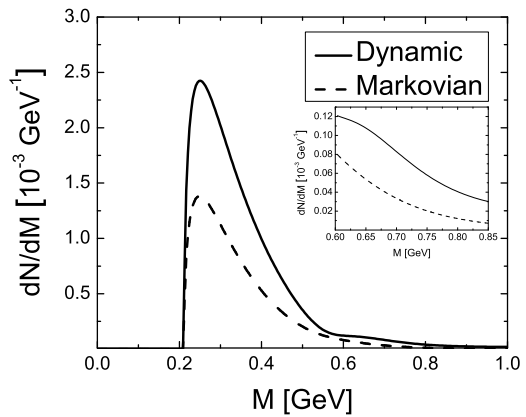}
        \caption{Dimuon yield from decaying $\rho$-mesons calculated out of equilibrium and within the Markovian approximation using parameterization (\ref{rapppar})
        for the meson mass and coupling to the $N^*(1520)$ resonance by means of the self energy (\ref{modelshenself}).}
        \label{fig:rappres}
      \end{center}
\end{figure}

The result of the calculation combining Brown-Rho scaling,
broadening and coupling to the $N^*(1520)$ is presented in Fig.
\ref{fig:rappres}. Here, the differences between the Markovian and
the dynamic calculations seem less significant as compared to the
mass shift scenarios with only marginal broadening (cf. Figs.
\ref{fig:rapp} and \ref{fig:br}). This can be expected since the
memory time scales approximately with the inverse width in the
system (see \cite{Schenke:2005ry}). The very broad spectral
function adjusts faster to the medium changes. However, around the
$\rho$-vacuum mass the dynamic calculation still leads to around
three times more yield than the Markovian one.

\begin{figure}[tbh]
      \begin{center}
        \includegraphics[height=5.5cm]{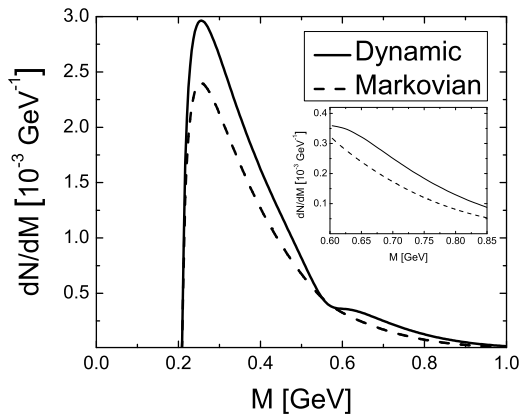}
        \caption{Dimuon yield from decaying $\rho$-mesons calculated out of equilibrium and within the Markovian approximation using coupling to the $N^*(1520)$ and broadening
        of both the vacuum and the resonance peak to 400 MeV. Maximal differences between the two calculations are about 30\%.}
        \label{fig:res}
      \end{center}
\end{figure}

For comparison we present the calculation for a scenario with
coupling to the resonance and broadening but without dropping mass
(Fig. \ref{fig:res}). The major difference to the scenario with
dropping mass is the overall increase of the yield. This is due to
the fact that in the dropping mass case, spectral weight is
shifted below the threshold of twice the muon mass at early times.
For the case without mass shift the difference between the dynamic
and Markovian calculations is less than 30\% for an invariant mass
up to 650 MeV (above that differences reach about 50\%) This is
again due to the very broad spectral function, for that memory
effects become less significant. Furthermore, the time scale for
adjustment to the medium \cite{Schenke:2005ry} also depends on the
pole mass of the spectral function. Lower masses mean larger
memory times, such that for dropping mass scenarios the effects
are naturally larger.

    In summary, we have shown that the inclusion of finite memory
    by calculation of the dilepton yield within a nonequilibrium
    field theoretical setup, using the realtime formalism, yields
    moderate to strong differences when comparing to dilepton
    yields, calculated assuming instantaneous adaption of all meson
    properties to the medium.
    Particularly, we have compared the results for dropping mass scenarios
    for dimuon yields in the NA60 scenario, which is modelled by an expanding fireball.
    An enhancement by about a factor of 3-4 around the $\rho$-vacuum mass and significant differences in shape have been
    found for the case of dropping mass with moderate broadening by 100 MeV. Inclusion of
    coupling to the $N^*(1520)$ resonance and significant broadening strongly modifies the shape of the calculated
    yields, especially in the dynamic calculation. However, the yield is still enhanced and more weight lies around the $\rho$-vacuum mass in the
    dynamic calculation. This is important for the comparison to the NA60 data, where the Markovian calculation using Brown-Rho scaling
    lead to too few dimuons in that region \cite{Brown:2005ka,Arnaldi:2006jq}.
    A scenario without mass shifts but including strong broadening \cite{vanHees:2006ng} is less affected by the inclusion of memory. The dynamic calculation
    differs from the Markovian one by maximally 50\%, and less than 30\% in most of the regarded range.

    We conclude that memory effects must not be neglected in
    precision calculations of dilepton yields from relativistic heavy ion
    collisions. Every dropping mass scenario we have investigated shows a significant dependence
    on whether instantaneous adaption to the medium or the full quantum
    mechanical evolution are considered.

\section*{Acknowledgments}
    B.S. thanks G.E. Brown and R. Rapp for stimulating discussions.
\bibliography{dileptons}
\end{document}